\documentclass[aps,showpacs,pra,twocolumn]{revtex4-1}

\usepackage{exscale}
\usepackage{graphicx}
\usepackage{amsmath}
\usepackage{latexsym}
\usepackage{epstopdf}
\usepackage{amsfonts}
\usepackage{amssymb}
\usepackage{bbm}
\usepackage{times}
\usepackage[T1]{fontenc}
\usepackage{lipsum}
\usepackage{amsthm}
\usepackage{fancyhdr,txfonts,bbm}
\usepackage{appendix}

\usepackage{hyperref}
\hypersetup{
    colorlinks=true,
    linkcolor=black,
    filecolor=black,      
    urlcolor=black,
    citecolor=black,
}
\usepackage{epsfig,pstricks}
\usepackage{changes}

\definechangesauthor[name={NKB}, color=orange]{NKB}

\newtheorem{definition}{Definition}

\newcommand{{\Cd}}{{\mathbb{C}^d}}
\newcommand{{\C}}{{\mathbb{C}}}

\newcommand{\Int}{\mathrm{int}}

\DeclareMathOperator{\Tr}{Tr}
\DeclareMathOperator{\atanh}{atanh}

\newcommand{\bra}[1]{\langle #1|}
\newcommand{\ket}[1]{|#1\rangle}

\newcommand{\ketbra}[2]{| #1\rangle\!\langle #2|}
\newcommand{\tr}[1]{\mbox{Tr}\left[ #1\right]}

\newcommand{\trA}[1]{\mbox{Tr}_A\left[ #1\right]}
\newcommand{\trB}[1]{\mbox{Tr}_B\left[ #1\right]}

\begin{document}

\title{A correlation measure detecting almost all non-Markovian evolutions}

%Markus Johansson$^1$}, and Antonio Ac\'\i n$^{1,3}$
\author{Dario De Santis$^1$, Markus Johansson$^1$, Bogna Bylicka$^1$, Nadja K. Bernardes$^{2,3}$, and Antonio Ac\'\i n$^{1,4}$}
\affiliation{$^1$ICFO-Institut de Ciencies Fotoniques, The Barcelona Institute of Science and Technology,
08860 Castelldefels (Barcelona), Spain\\
$^2$ Departamento  de  F\'\i sica,  Universidade  Federal  de  Minas  Gerais,
Belo  Horizonte,  Caixa  Postal  702, 30161-970, Brazil\\
$^3$ Departamento de F\'\i sica, Universidade Federal de Pernambuco, 50670-901 Recife-PE, Brazil\\ 
$^4$ICREA--Institucio Catalana de Recerca i
Estudis Avan\c{c}ats, Lluis Companys 23, 08010 Barcelona, Spain}

\date{\today}
\begin{abstract}
We investigate the ability of correlation measures to witness non-Markovian open quantum system dynamics.
It is shown that the mutual information and any entanglement measure between the system and an ancilla do not witness all non-Markovian dynamics. A correlation measure is introduced, and it is proven that, in an enlarged setting with two ancillary systems, this measure detects almost all non-Markovian dynamics, except possibly a zero-measure set of dynamics that is non-bijective in finite time-intervals. Our proof is constructive and provides different initial states detecting the non-Markovian evolutions. These states are all separable and some are arbitrarily close to a product state. 
\end{abstract}

\maketitle

\section{Introduction}
The dynamics of open quantum systems \cite{book_B&P, book_Weiss, book_R&H} has been investigated extensively in recent years for both fundamental and applicative reasons.
In particular the phenomenon of reservoir memory effects has been studied since such effects can induce a recovery of correlations or coherence and are therefore viewed as a potential resource for the performance of quantum technologies.
The problem of characterising memoryless dynamics, the so-called Markovian regime, and dynamics exhibiting memory effects, the non-Markovian regime, has been considered in a wide range of different ways (for extended reviews see \cite{rev_RHP, rev_BLP}). While a unique agreed upon concept of quantum Markovianity does not exist, it is frequently identified with the property of { \it Completely Positive divisibility} (CP-divisibility).
An evolution is CP-divisible if between any two points in time it can be described by a CP-map.  
This idea generalises the semigroup property \cite{GKS, L} of classical Markovian processes. 
{In this work, we follow this convention and adopt CP-divisibility as formal definition of Markovianity.} 

A complementary way of addressing {memory effects} consists of identifying operational quantities that can detect the information backflow expected in non-Markovian evolutions~\cite{RHP,Luo,Buscemi&Datta,bogna, rivas}.
A common approach is to study functions that are monotonically non-increasing under local CP-maps.
An increase of such a quantity implies that the evolution is not CP-divisible, {hence non-Markovian}, although the converse may not be true in general. Investigating under what conditions a non-increase of these quantities is in one-to-one correspondence with CP-divisibility is relevant for evaluating current methods for non-Markovianity detection, finding new ones, and understand the operational consequences of non-Markovianity.
It is also relevant to understand how these different detection methods are related, and to what extent they are equivalent.
In particular, it has been shown that the guessing probability of minimum error state discrimination can be used to witness any non-Markovian dynamics \cite{Buscemi&Datta}. However, no method for constructing state ensembles required for this is known. A constructive method to witness any bijective non-Markovian dynamics using an ensemble of two equiprobable states has subsequently been proposed \cite{bogna}. 

In this work  we investigate the relation between non-Markovianity and correlations.
We first show that the quantum mutual information between system and ancilla as well as any entanglement measure are unable to witness all non-{Markovian} dynamics.
The next natural question is to understand whether there exist such correlation measures. To investigate this, we first introduce a bipartite correlation measure based on the distinguishability of an ensemble of remotely prepared states. We then use this measure in an extended setting consisting of the system and two ancillary systems and prove that the non-increase of this measure is in one-to-one correspondence with CP-divisibility for almost all evolutions. More precisely, we show how to detect a correlation backflow for all non CP-divisible evolutions that are bijective or at most point-wise non-bijective. Our method is constructive and provides a family of initial states able to detect the correlation backflow. The states in this family are all separable and include states that are arbitrarily close to uncorrelated.

\section{Non-Markovian dynamics}
We denote the set of bounded operators on any finite dimensional Hilbert space $\mathcal{H}$ by $B(\mathcal{H})$ and that of positive semidefinite trace-one operators, i.e., the set of quantum states, by $S(\mathcal{H})$. In the following, we consider two finite dimensional systems: a quantum system $S$ with a Hilbert space $\mathcal{H}_S$ and an ancillary system $A$ with a Hilbert space $\mathcal{H}_A$.

The evolution of $S$ from initial time $0$ to a later time $t$ is described by a dynamical map, i.e., a {completely positive and trace preserving (CPTP) linear operator $\Lambda_t: B(\mathcal{H}_S)\rightarrow B(\mathcal{H}_S)$}. The dynamics of the system is thus described by the family of maps $\{\Lambda_t\}_t$ parametrized by $t$.
An important concept for the study of non-Markovian effects is the divisibility of the dynamical map, as well as Positive (P) and Completely Positive (CP) -divisibility in terms of intermediate maps $V_{s,t}$.

\begin{definition}
A dynamical map $\Lambda_s$ is called (P/CP) divisible if it can
be expressed as a sequence of linear trace preserving (P/CP) maps $\Lambda_s=V_{s,t}\Lambda_t$, where $V_{s,t}$ is a linear trace preserving (P/CP) map, for any $0\leq t\leq s$.
\end{definition}
{As said}, CP-divisibility is a common definition of Markovian evolution~\cite{book_R&H} {that we also adopt in this work}.

\section{Correlation measures that are insufficient as witnesses}
We start by showing the limitations of the ordinarily used correlation measures in identifying all non-Markovian dynamics. In particular, we consider correlation measures between system $S$ and an ancilla $A$ where $\Lambda_t$ acts only on $S$.

A correlation measure is a function $M$ such that, (i) $M\geq 0$; (ii) $M(\rho)=0$ if $\rho$ is a product state; (iii) $M$ is non-increasing under local operations.
Condition (iii) implies that all correlation measures are non-increasing for local CP-divisible evolutions.
Thus, if an increase in correlation occurs between time $t$ and $s$ there is no CP intermediate map $V_{s,t}$. This property explains why correlation measures have been utilized to witness and quantify non-Markovian effects~\cite{RHP,Luo}.

\subsection{Entanglement measures}
An entanglement measure $M_E$ \cite{vidal} is a correlation measure that satisfies the additional condition of non-increase under local operations aided by classical communication. This implies that  $M_E(\rho)=0$ if $\rho$ is a separable state. The idea of using an entanglement measure to witness non-Markovianity was first introduced in Ref. \cite{RHP}. However, for any entanglement measure there are non CP-divisible dynamics that cannot be witnessed. 
Consider for instance an evolution that consists first of an entanglement breaking \cite{horo} dynamical map $\Lambda_t$ that maps any state to a separable state, e.g. a sufficiently depolarizing map. 
Any entanglement measure is zero everywhere on the image of such an evolution.
If the dynamics following this entanglement breaking evolution $\Lambda_t$ is non-Markovian P-divisible, separable states are mapped to separable states. Then any entanglement measure is non-increasing, because it remains equal to zero, and thus fails to detect the non-Markovianity.

\subsection{The mutual information}
Another commonly used correlation measure is the 
quantum mutual information $I(\rho)$ \cite{straton}. For states $\rho_{AS}\in S(\mathcal{H}_A\otimes \mathcal{H}_S)$ it is defined as 
\begin{eqnarray}
I(\rho_{AS})\equiv S(\rho_A)+S(\rho_S)-S(\rho_{AS}),
\end{eqnarray}
where $S(\rho)$ is the von Neumann entropy of the state $\rho$, and $\rho_A,\, \rho_S$ are the reduced states of $A$ and $S$, respectively. A measure of non-Markovian effects based on the mutual information was proposed in Ref. \cite{Luo}. {Since the mutual information is non-increasing under CP maps, an increase of this quantity is sufficient for detecting non-Markovianity.} 
However, {it is not necessary:} we demonstrate {in what follows} {the existence of non-Markovian evolutions} for which no increase in the mutual information occurs. 

{To construct this example, we consider} random unitary dynamics, defined as convex combinations of unitary dynamics. For a qubit, these can be represented by the dynamical maps
\begin{eqnarray}\label{ranu}
\Lambda_{t}(\sigma_x)&&=e^{-\int_{0}^{t}(\gamma_z(\tau)+\gamma_y(\tau))d\tau}\sigma_x,\phantom{u}
\Lambda_{t}(\sigma_y)=e^{-\int_{0}^{t}(\gamma_z(\tau)+\gamma_x(\tau))d\tau}\sigma_y,\nonumber\\
\Lambda_{t}(\sigma_z)&&=e^{-\int_{0}^{t}(\gamma_x(\tau)+\gamma_y(\tau))d\tau}\sigma_z,\phantom{u}
\Lambda_{t}(\mathbbm{1}_S)=\mathbbm{1}_S,
\end{eqnarray}
where $\gamma_k(t)$ are real valued functions of $t$, $\sigma_k\in B(\mathcal{H}_S)$ for $k=x,y,z$ are the Pauli matrices and $\mathbbm{1}_S$ is the identity opertor on $\mathcal{H}_S$.
The dynamics is CP-divisible if and only if $\gamma_k(t)\geq 0$ for $k=x,y,z$ \cite{book_R&H,darekk}, and P-divisible if and only if 
$\gamma_i(t)+\gamma_j(t)\geq{0}$ for all $i\neq j$ since the intermediate map $V_{s,t}$ is then contractive in the trace norm \cite{darekk,koss1,koss2,ruskai}. 
The stationary states of the dynamics are $\rho_A\otimes \mathbbm{1}_S/2$ where $\rho_A$ is any state in $S(\mathcal{H}_A)$.

We can introduce an orthonormal basis $\{e_i\}_i$ of $B(\mathcal{H}_A\otimes\mathcal{H}_S)$ with corresponding coordinates $\bar{a}\equiv \{a_i\}_i$, i.e., $\Tr(e_i  e_j)=\delta_{ij}$ and $a_i=\Tr(\rho e_i)$.
Then, as we show in Appendix \ref{meth}, if $\gamma_k(t)$ are continuous functions of $t$, the time derivative $\frac{d}{dt}I(\bar{a},t)$ as a function of $\bar{a}$ is analytic in the interior of $S(\mathcal{H}_A\otimes\mathcal{H}_S)$.
Where it is analytic $\frac{d}{dt}I(\bar{a},t)$ can be described in a neighbourhood of the stationary states by Taylor expansions in $\bar{a}$. 

By definition, $\frac{d}{dt}I(\bar{a},t)=0$ at any stationary state.
For the case of $\dim{\mathcal{H}_A}=2$, the first and second derivatives, $\frac{\partial}{\partial a_i}\frac{d}{dt}I(\bar{a},t)$ and $\frac{\partial^2}{\partial a_i\partial a_j}\frac{d}{dt}I(\bar{a},t)$, were calculated at the stationary states in the interior of $S(\mathcal{H}_A\otimes \mathcal{H}_S)$, using a method adapted from \cite{Tsing} (see Appendices \ref{Taylor} and \ref{mutual}). All first derivatives are identically zero.
From the second derivatives the eigenvalues of the Hessian matrix were obtained. Each non-zero eigenvalue is proportional to $[\gamma_i(t) + \gamma_j(t)]$ for some $i\neq j$, and is non-positive for P-divisible dynamics and negative if $\gamma_i(t) + \gamma_j(t)>0$. On the zero eigenspace of the Hessian and in the neighbourhood of the stationary states in the boundary of $S(\mathcal{H}_A\otimes \mathcal{H}_S)$ we directly evaluated $\frac{d}{dt}I(\bar{a},t)$. On the zero-eigenspace $\frac{d}{dt}I(\bar{a},t)$ is non-positive for P-divisible dynamics, and the neighbourhood of the stationary states in the boundary of $S(\mathcal{H}_A\otimes \mathcal{H}_S)$ contains only product states where $\frac{d}{dt}I(\bar{a},t)=0$ (see Appendix \ref{mutual}).

In conclusion there exist non-Markovian  dynamics for which there is a neighbourhood of the stationary states where the time derivative of the mutual information is non-positive. Moreover, for the considered dynamics, it is always possible to tune the rates $\gamma_k(\tau)$ for $0\leq\tau\leq t$ so that the non-Markovian regime starts only after a time $t$ for which the image of $\Lambda_t$ is contained in this neighbourhood of the stationary states with non-increasing mutual information (see Appendix \ref{tuning}). Therefore, there exist evolutions for which the non-Markovian character cannot be witnessed by an increase in the mutual information.

\section{Introduction of a correlation measure}

We have seen how ordinarily used correlations measures fail to detect many cases of non-Markovian dynamics. Now, one may wonder if this limitation applies to any correlation measure or, on the contrary, if there exists a correlation measure that witnesses any non-Markovian dynamics. Motivated by this question we introduce a correlation measure based on the distinguishability of the ensembles one party prepares for the other party by performing local measurements on half of a bipartite state. {For this, we first} need to discuss several concepts related to the distinguishability of quantum states.

\subsection{Maximally entropic measurements}
Consider a system in a quantum state $\rho$. An $n$-outcome measurement on this system is represented by a positive-operator valued measure (POVM), i.e., a collection of positive semi-definite operators  $\{ P_{i}\}_{i=1}^n$ {such that $\sum_{i=1}^{n} P_{i} = \mathbbm{1}$}. Each $ P_{i}$ represents a possible outcome with the probability of occurrence {$p_i=\tr{\rho P_i}$}. 

We say that a POVM  is \textit{maximally entropic} (ME-POVM)  \textit{for} $\rho$ if, when applied on $\rho$, each outcome has the same probability of occurrence: $p_i=1/n$. Indeed,  if  $S(\left\{ p_i \right\}_i)=-\sum_i p_i \log_n p_i$ is the Shannon entropy of the resulting $n$-outcome probability distribution, where we take as the basis of the logarithm in the entropy the number of outputs, $S(\{p_i\}_i)=1$ if and only if $p_i = 1/n$. We define the set of ME-POVMs for $\rho$ as
\begin{equation}\label{MEPOVM}
\Pi \left( \rho \right) \equiv \left\{ \left\{ P_{i} \right\}_{i} : S(\left\{ p_i \right\}_i)=1  \right\} \, .
\end{equation}
For any state $\rho$, this collection is non-empty and contains measurements with any number of outputs (see Appendix \ref{neverempty}).

\subsection{Guessing probability of an ensemble}
Consider an ensemble of states $\mathcal{E} = \{ p_i , \, \rho_{i} \}_{i=1}^n$. The average probability to correctly identify a state extracted from $\mathcal{E}$, maximized over all possible measurements, is called the \textit{guessing probability} of the ensemble
\begin{equation}\label{Pg}
P_g (\mathcal{E})\equiv \max_{ \left\{ P_i \right\}_i }\sum_{i=1}^n  p_i \tr{ \rho_i \, P_i  } \, ,
\end{equation}
where the maximization is performed over the space of the $n$-output POVMs. Using the definition $\overline{p} \equiv \max_i \{p_i\}_{i=1}^n \geq 1/n$, it follows that $P_g( \mathcal{E}) \geq \overline{p}$ , where the equality holds if $\mathcal{E}$ is made of identical states. Hence, {$P_g(\{ p_i=1/n, \rho_i=\rho \}_{i=1}^n) =1/n$}.

Note that when the ensemble is composed by two equiprobable states, i.e. $ \mathcal{E}^{eq}=\{\{p_{1,2}=1/2\},\{\rho_1, \, \rho_2\}\}$, $P_g(\mathcal{E}^{eq})$ can be expressed in terms of the distinguishability {$D(\rho_1,\rho_2)\equiv||\rho_1 - \rho_2 ||_1 /2$ between $\rho_1$ and $\rho_2$
\begin{equation}\label{PgD}
P_g(\mathcal{E}^{eq} ) = \frac{1}{4} \left( 2 + ||\rho_1 - \rho_2 ||_1 \right) \, ,
\end{equation}
where $|| \cdot ||_1$ is the trace norm.}

\subsection{Definition of the correlation measure}
We now have all the ingredients needed to define our correlation measure. Consider a bipartite state $\rho_{AB}$ defined on a finite dimensional state space of a composed system $S(\mathcal{H}_A \otimes \mathcal{H}_B)$. A measurement with an arbitrary number of outcomes $\{P_{A,i} \}_i$ performed on system $A$ prepares on $B$ the ensemble $\mathcal{E}(\rho_{AB}, \{P_{A,i}\}_i ) \equiv \{ p_i , \rho_{B,i}\}_i$ defined by
\begin{equation}\label{prhooutputs}
p_i= \tr{ \rho_{A} \, P_{A,i} } , \, \rho_{B,i} = \frac{\trA{ \rho_{AB}  \, P_{A,i} \otimes \mathbbm{1}_B }}{ p_i }\, ,
\end{equation}
where $\rho_A=\trB{\rho_{AB}}$ is the reduced state on $A$. From Eq. (\ref{prhooutputs}) it follows that $\{ \{P_{A,i} \}_i: \{P_{A,i} \otimes \mathbbm{1}_B\}_i \in \Pi (\rho_{AB})\} = \Pi(\rho_A)$.

A correlation measure $C_A^{(2)}$ is obtained by maximizing the guessing probability of these ensembles over  the 2-output ME-POVMs on $A$,
\begin{equation}\label{CA}
 C_A^{(2)} (\rho_{AB}) \equiv \max_{ \left\{ P_{A,1}, P_{A,2}\right\} \in \Pi \left( \rho_{A} \right) }  P_g \left( \mathcal{E} \left( \rho_{AB} ,\left\{ P_{A,1}, P_{A,2}\right\}  \right) \right)  - \frac{1}{2} \, . 
\end{equation}
Alternatively, we could perform 2-output ME-POVMs on the system $B$ and obtain a measure
\begin{equation}\label{CB}
 C_B^{(2)} (\rho_{AB}) \equiv \max_{ \left\{ P_{B,1}, P_{B,2}\right\} \in \Pi \left( \rho_{B} \right) }  P_g \left( \mathcal{E} \left( \rho_{AB} ,\left\{ P_{B,1}, P_{B,2}\right\}  \right) \right)  - \frac{1}{2} \, ,
\end{equation}
where $\rho_B=\trA{\rho_{AB}}$ is the reduced state on $B$.
We underline that the guessing probabilities that appear in Eq.~(\ref{CA}) and (\ref{CB}) can be evaluated using Eq. (\ref{PgD}).
A natural way to  construct a symmetric measure with respect to $A$ and $B$ is the following
\begin{equation}\label{C}
 C^{(2)} (\rho_{AB}) \equiv \max \, \left\{ C^{(2)}_A(\rho_{AB}), \, C^{(2)}_B(\rho_{AB}) \right\}  \, . 
\end{equation}
Operationally, $C^{(2)}_A (\rho_{AB})$ ($C^{(2)}_B (\rho_{AB})$) corresponds to the largest distinguishability between the pairs of equiprobable states of $B$ ($A$) that we can obtain from $\rho_{AB}$ by performing measurements on $A$ ($B$).

Similar correlation measures $C^{(n)}$ can be obtained by fixing the number of outputs of the ME-POVMs to any integer $n\geq 3$ and replacing the term $1/2$ in Eq. \eqref{CA} and \eqref{CB} by $1/n$. Moreover, we define $ C(\rho_{AB}) \equiv \max \, \{ C_A (\rho_{AB}), \, C_B (\rho_{AB}) \} $, where $C_A(\rho_{AB})$ ($C_B(\rho_{AB})$) is obtained without fixing the number of outputs of the ME-POVMs in $\Pi(\rho_A)$ ($\Pi(\rho_B)$), namely
\begin{equation}\label{C_Aorig}
C_A (\rho_{AB}) \equiv \max_{ \{P_{A,i}  \}_i \in \Pi \left( \rho_{A} \right) }  P_g \left( \mathcal{E} \left( \rho_{AB} ,\left\{ P_{A,i}\right\}_i  \right) \right)  - \frac{1}{2} \, . 
\end{equation} 
We define $C_B(\rho_{AB})$ similarly.

To show that $C$ and $C^{(n)}$, for any $n\geq 2$, are proper correlation measures, we must prove that they are: (i) non-negative, (ii) zero-valued for product states and (iii) monotone under local operations. 
First, we prove that property (ii)  holds for $C^{(2)}_A$.
For any product state $\rho_{AB}=\rho_A\otimes\rho_B$ and ME-POVM on $A$, the equiprobable output states $\rho_{B,1}$ and $\rho_{B,2}$ are identical. Hence, $\rho_{B,1}=\rho_{B,2}=\rho_B$ and $C^{(2)}_A( \rho_A\otimes\rho_B )=||\rho_{B,1}-\rho_{B,2}||/4=0$. The generalizations to prove that (ii) is valid {also} for $C^{(n)}$ for any {$n\geq 2$} and $C$ are obvious. Consequently,  property (i) is trivial. 
In Appendix \ref{monolocal} we prove that the monotonicity property (iii) holds for $C$ and $C^{(n)}$, for any $n\geq 2$, and therefore they are proper correlation measures.

While we have defined a whole class of correlation measures, in the following we focus on the potential of $C^{(2)}$ to witness non-Markovian dynamics. Therefore, unless otherwise specified the correlation measure referred to is $C^{(2)}$.

\section{Witnessing non-Markovian dynamics}
\begin{figure}
\includegraphics[width=0.45\textwidth]{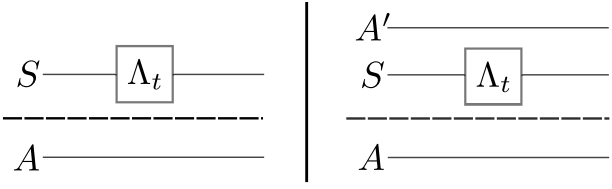}
\caption{Left: in the standard setting, an initial state between system $S$ and ancilla $A$ is used. An increase of correlations between these two parts witnesses the presence of non-Markovian effects. Right: in our extended setting, the whole system consists of three parts, system $S$ and ancilla $A$ as before, plus an extra ancilla $A'$. An increase of the correlations over the bipartition $A$ versus $SA'$ is used to witness non-Markovian evolutions.}\label{fig}
\end{figure}
We now show how to use the correlation measure introduced above to detect non-Markovian evolutions. We prove that for any evolution that is at most point-wise non-bijective, we can find an initial state $\rho_{AB}^{(\tau)}(0)$ such that $C^{(2)} (\rho_{AB}^{(\tau)}(t))$ increases between time $t=\tau$ and $t=\tau+ \Delta t$ if and only if there is no CP intermediate map $V_{\tau+\Delta t,\tau}$.
By at most point-wise non-bijective evolutions we refer to evolutions where multiple initial states are mapped to the same state by $\Lambda_t$ for at most a discrete set of times $t$.
 Although our method applies to any bijective or pointwise non-bijective evolution, at the moment we are unable to extend the proof to non-Markovian evolutions that are non-bijective in finite time intervals. Note however that the set of non-Markovian evolutions not covered by our result has zero measure in the space of evolutions. More precisely, if we take an evolution that is non-bijective in a finite time interval and add a perturbation chosen at random with respect to a Borel measure, this yields an at most point-wise non-bijective evolution with probability one \cite{ottyorke}.  

To take full advantage of this measure, we extend the standard setting and consider a scenario where $A$ is an ancillary qubit and $B$ is composed of the system $S$ undergoing evolution and a suitably chosen ancilla $A'$, see Fig.\ref{fig}. First, we construct the state $\rho_{AB}^{(\tau)}(t)$ to be used as a probe. Second, we show that for the class of non-Markovian dynamics specified above, $C^{(2)} ( \rho_{AB}^{(\tau)}(t))$ provides a correlation backflow.

\subsection{The probe}
Let $\Lambda_t$ represent a bijective or pointwise non-bijective non-Markovian dynamical map that acts on the system $S$ and introduce an ancillary system $A'$.
As shown in Ref. \cite{bogna}, for any of these dynamics we can construct a class of pairs of initial states $\{ \rho_B'^{(\tau)}(0), \rho_B''^{(\tau)}(0)\}\in S(\mathcal{H}_B )= S (\mathcal{H}_{A'} \otimes \mathcal{H}_{S})$ that show an increase in distinguishability between time $t=\tau$ and $t=\tau +\Delta t$
\begin{equation}\label{rho12}
\big| \big|  \rho_B'^{(\tau)}(\tau +\Delta t) - \rho_B''^{(\tau)}(\tau +\Delta t)\big| \big|_1 >\big| \big|  \rho_B'^{(\tau)}(\tau ) - \rho_B''^{(\tau)}(\tau )\big| \big|_1  \, ,
\end{equation}
if and only if there is no CP intermediate map $V_{\tau+\Delta t, \tau}$, where the evolution of the system $B$ is given by the dynamical map $\mathcal{I}_{A'}\otimes \Lambda_{t}$, where $\mathcal{I}_{A'}$ is the identity map on $A'$.

The particular bipartite separable states $\rho_{AB}^{(\tau)}(t)$ for which we examine the correlation $C^{(2)}$ are classical-quantum states%\cite{wilde}
. Our ``probe'' state is
\begin{equation}\label{hatrho}
\rho_{AB}^{(\tau)}(t) \equiv \frac{1}{2} \left( \ketbra{0}{0}_A \otimes \rho_B'^{(\tau)} (t)+ \ketbra{1}{1}_A \otimes \rho_B''^{(\tau)} (t) \right) \, ,
\end{equation}
where and $ \rho_B'^{(\tau)}(t)$ and $ \rho_B''^{(\tau)}(t) $ are the states that appear in Eq. (\ref{rho12}) and $\mathcal{B}_A \equiv\{\ket{0}_A, \ket{1}_A \}$ is an orthonormal basis for $\mathcal{H}_A$. Since only the system $B$ is involved in the evolution, $\rho_{AB}^{(\tau)}(t)$ is given by Eq. (\ref{hatrho}) for any $t\geq 0$. Note that from Eq. (\ref{hatrho}) it follows that $\rho_{AB}^{(\tau)}(t)$ does not contain any entanglement. Moreover, the state can be chosen arbitrarily close to an uncorrelated state since, as shown in \cite{bogna}, one can always choose states $\rho_B'^{(\tau)}(0)$ and $\rho_B''^{(\tau)}(0)$ arbitrarily close to each other.

\subsection{Detecting the correlation backflow}
 We now show how the correlation measure $C^{(2)}_A(\rho_{AB}^{(\tau)}(t))$, and later $C^{(2)}(\rho_{AB}^{(\tau)}(t))$, witnesses bijective {or pointwise non-bijective} non-Markovian dynamics. 

To evaluate $C^{(2)}_A(\rho_{AB}^{(\tau)}(t))$, we have to find a ME-POVM $\{P_{A,1},P_{A,2}\}$ that, applied on $\rho_{AB}^{(\tau)}(t)$, generates the output ensemble $\{ \{p_{1,2}=1/2\},\{\rho_{B,1}(t),\rho_{B,2}(t)\}\}$ with the largest value of $||\rho_{B,1}(t)- \rho_{B,2}(t) ||_1$. Let $\lambda \in[0,1]$  and $\eta\in[0,1]$  be the diagonal elements of $P_{A,1}$ in the basis $  \mathcal{B}_A$. It is easy to show that $\lambda+\eta=1$ for ME-POVMs. The corresponding output states are
\begin{eqnarray}
\rho_{B,1}(t)&=&{\lambda \rho_{B}'^{(\tau)}(t) + \eta \rho_B''^{(\tau)}(t)} \, , \\
\rho_{B,2}(t)&=&{(1-\lambda) \rho_{B}'^{(\tau)}(t) +(1-\eta) \rho_B''^{(\tau)}(t)} \, .
\end{eqnarray}
It follows that
\begin{equation}
||\rho_{B,1}(t)- \rho_{B,2} (t) ||_1 =  {|\lambda-\eta| \cdot ||\rho_{B}'^{(\tau)}(t) - \rho_B''^{(\tau)}(t)||_1 }    \, . 
\end{equation}
Since $0\leq |\lambda-\eta| \leq 1$, the maximum is obtained when either $\lambda$ or $\eta$ is equal to 1. In both cases the output states are $\rho_B'^{(\tau)}(t)$ and $\rho_B''^{(\tau)}(t)$ and 
\begin{equation}\label{CAhatrho2}
C^{(2)}_A( \rho_{AB}^{(\tau)}(t)) =  \frac{|| \rho_B'^{(\tau)} (t) - \rho_B''^{(\tau)}(t)||_1}{4}   \, .
\end{equation}
In Appendices \ref{CB2CA} and \ref{2enoughA} we prove that $C^{(2)} (\rho_{AB}^{(\tau)}(t)) = C^{(2)}_A(\rho_{AB}^{(\tau)} (t) ) \geq {C_B^{(2)} (\rho_{AB}^{(\tau)} (t) )}$.
Therefore, using Eqs. (\ref{rho12}) and (\ref{CAhatrho2}), we obtain a correlation backflow
\begin{equation}\label{ddtCA}
C^{(2)}\left(\rho_{AB}^{(\tau)}(\tau+\Delta t) \right) > C^{(2)}\left(\rho_{AB}^{(\tau)}(\tau) \right)  \, ,
\end{equation}
if and only if there is no CP intermediate map $V_{\tau+\Delta t,\tau}$. 

In Appendices \ref{2enough} and \ref{2enoughA} we prove that $C_B(\rho_{AB}^{(\tau)}( t))=C_B^{(2)}(\rho_{AB}^{(\tau)}( t))$ and $C_A(\rho_{AB}^{(\tau)}(t)) = C_A^{(2)}(\rho_{AB}^{(\tau)}(t))$. From these additional results it follows that for this initial probe state $C^{(2)} (\rho_{AB}^{(\tau)}( t)) = C (\rho_{AB}^{(\tau)}( t)) $ at any time $t\geq 0$.

\section{Discussion}
The main motivation of this work is to understand the power of correlations to witness non-Markovian evolutions. We have first provided examples of  non-Markovian random unitary qubit evolutions for which the quantum mutual information between system and ancilla never increases. Moreover, we have pointed out that any entanglement measure is insufficient for witnessing any P-divisible non-Markovian dynamics that takes place after an initial Markovian entanglement breaking evolution. 
We then introduced a correlation measure and showed that, in an extended setting with a second ancilla, it displays backflow for almost all non-Markovian evolutions. More precisely, it displays backflows for all non-Markovian evolutions that are bijective or at most point-wise non-bijective. For a given dynamics we described how states that exhibit such an increase in correlations can be constructed. These states have no entanglement across the given bipartition and can be chosen to be arbitrarily close to an uncorrelated state.

The question if there exists a measure of correlation with the property of being non-increasing if and only if the dynamics is CP-divisible, without any restrictions on the dynamics, is still open, both in the case of system-ancilla correlations and in the extended setting with a second ancilla. A possible avenue consists of understanding how to adapt the results in~\cite{Buscemi&Datta}, valid for any non-Markovian evolution, to our correlation measure. Another open question is to understand if the use of the second ancilla provides an advantage for other correlation measures, as it happened for the correlation measure considered in this work.

\begin{acknowledgments}

Support from the ERC CoG QITBOX, the AXA Chair in Quantum Information Science,
Spanish MINECO (QIBEQI FIS2016-80773-P and Severo Ochoa SEV-2015-0522),
Fundaci\'o Privada Cellex, and the
Generalitat de Catalunya (CERCA Program and SGR1381) is acknowledged.
% Spanish MINECO (FOQUS FIS2013-46768-P and SEV-2015-0522), Fundacion Cellex, and the Generalitat de Catalunya (SGR 875and CERCA Program) is acknowledged.
This project has received funding from the European Union's Horizon 2020 research and innovation programme under the Marie Sk\l odowska-Curie grant agreement No 665884.
D.D.S. acknowledges support from the ICFOstepstone programme, funded by the Marie Sk\l odowska-Curie COFUND action (GA665884).
B.B. acknowledges support from an ICFO-MPQ
Fellowship. N.K.B. thanks the Brazilian funding agency CAPES.

\end{acknowledgments}

\appendix

\section{Generators of differentiable evolutions}

For a differentiable evolution, any dynamical map $\Lambda_t$, and any intermediate map $V_{s,t}$, can be expressed as a time ordered exponential
\begin{eqnarray}
\Lambda_t=\mathcal{T}e^{\int_{0}^{t}\mathcal{L}_{\tau}d\tau},\phantom{uu} V_{s,t}=\mathcal{T}e^{\int_{t}^{s}\mathcal{L}_{\tau}d\tau},
\end{eqnarray}
where $\mathcal{L}_{t}$ is the Gorini-Kossakowski-Sudarshan-Lindblad generator \cite{GKS,L,book_R&H} of the evolution, defined as
\begin{eqnarray}\label{geno}
\mathcal{L}_t(\rho)\equiv i[H(t),\rho]+\sum_k\gamma_{k}(t)\left( G_k(t)  \rho G_k^{\dagger}(t)-\frac{1}{2}\left\{G_k^{\dagger}(t) G_k(t),\rho\right\}\right), \nonumber \\ 
\end{eqnarray}
where the $\gamma_k(t)$ are real time dependent functions, the $G_k(t)$ are time dependent operators and $H(t)$ is a Hermitian time dependent operator. 
The generator $\mathcal{L}_t$ gives rise to Markovian evolution if and only if it can be written on a form where $\gamma_k(t)\geq 0$ for all $k$ (see e.g. \cite{book_R&H}).

\section{Taylor expansion of the time derivative of the mutual information}\label{meth}

As a tool to investigate the time dependence of the mutual information we delineate how the time derivative of the of the mutual information $\frac{d}{dt}I(\bar{a},t)\equiv\frac{d}{ds}I[\bar{a},V_{s,t}]\big{|}_{s=t}$ can be described by a Taylor expansion in the $a_i$ at $\bar{a}$. In particular we consider its behaviour in neighbourhoods of the stationary states of a given evolution.

The mutual information $I(\rho)$ is analytic for all $\rho$ of full rank, i.e., everywhere in the interior $\Int[S(\mathcal{H}_A\otimes \mathcal{H}_S)]$ of the set of states. Thus, for any  open neighbourhood $U\subset \Int[S(\mathcal{H}_A\otimes \mathcal{H}_S)]$ of $\rho$ the mutual information equals its Taylor series and we can use Taylor expansions to analyse its local properties. Moreover, the time derivative of the mutual information is also analytic if the dynamics is differentiable.
To see this note that the time derivative $\frac{d}{dt}I(\bar{a},t)\equiv\frac{d}{ds}I[\bar{a}, V_{s,t}  ]\big{|}_{s=t}$ can be expressed as $\frac{d}{dt}I(\bar{a},t)=\sum_{i,j}a_j\frac{d V_{ij}(s,t)}{ds}\big{|}_{s=t}\frac{\partial}{\partial a_i}I(\bar{a},t)$ where $V_{ij}(s,t)\equiv \Tr[e_i\mathcal{I}_A\otimes V_{s,t}(e_j)]$ .
Next, assume that $\frac{d V_{ij}(s,t)}{ds}\big{|}_{s=t}$ is well defined for each $ij$. Then,
since products, linear combinations, and derivatives of analytic functions are analytic it follows that $\frac{d}{dt}I(\bar{a},t)$ is analytic as a function of $\bar{a}$ in $\Int[S(\mathcal{H}_A\otimes \mathcal{H}_S)]$.

Thus, if $V_{s,t}$ is differentiable $\frac{d}{dt}I(\bar{a},t)$ can be described in an open neighbourhood of any state in $\Int[S(\mathcal{H}_A\otimes \mathcal{H}_S)]$ by its Taylor expansion. On the boundary of $S(\mathcal{H}_A\otimes \mathcal{H}_S)$ on the other hand the partial derivatives in $\bar{a}$ need not even be well defined to all orders.

Let $\bar{a}_0$ be the coordinate of a stationary state in $\Int[S(\mathcal{H}_A\otimes \mathcal{H}_S)]$ of a linear divisible dynamic described by $\Lambda_t$.
Since $\frac{d}{dt}I(\bar{a}_0,t)=0$ the sign of $\frac{d}{dt}I(\bar{a},t)$ in a neighbourhood of $\bar{a}_0$ is determined by the terms of higher order than zero of the Taylor expansion of $\frac{d}{dt}I(\bar{a},t)$ with respect to $\bar{a}$.

\subsection{Neighbourhoods of critical points}
 
Unless all first derivatives are non-zero it is necessary to consider higher order terms of the Taylor expansion. In particular this is true if all first derivatives with respect to $\bar{a}$ are zero, i.e., if $\bar{a}_0$ is a critical point of $\frac{d}{dt}I(\bar{a},t)$.

The nature of a critical point $\bar{a}_0$ can be investigated by calculating the eigenvalues of the Hessian matrix, i.e., the matrix $\textbf{H}_{i,j}=\frac{\partial^2}{\partial a_i \partial a_j}\frac{d}{dt}I(\bar{a},t)$. 
However, at a stationary state, the Hessian $\textbf{H}_{i,j}$ does not have full rank since $\frac{d}{dt}I(\bar{a},t)=0$ on the set of stationary states $S_s$ of $V_{s,t}$, and on all product states $S_p$.
Therefore any eigenvector of the Hessian that is tangent to $S_s\cup S_p$ corresponds to a zero eigenvalue. 
The behaviour of $\frac{d}{dt}I(\bar{a},t)$ on the zero-eigenspace $E_0$ of  $\textbf{H}_{i,j}$ cannot be determined from the Hessian matrix since it depends on higher order derivatives.

On the complement of $E_0$, i.e., on $E_0^C\equiv B(\mathcal{H}_A\otimes \mathcal{H}_S)\backslash E_0$, the Hessian does describe the behaviour of $\frac{d}{dt}I(\bar{a},t)$ in some neighbourhood of $\bar{a}_0$.
In particular, if all eigenvalues of the Hessian that correspond to eigenvectors tangent to $E_0^C$ are negative there exist some neighbourhood $U^-_{\bar{a}_0}$ of $\bar{a}_0$ where $\frac{d}{dt}I(\bar{a},t)$ is negative in $U^-_{\bar{a}_0}\cap E_0^C$. If all eigenvalues of the Hessian that correspond to eigenvectors that are tangent to $E_0^C$ are positive there exist some neighbourhood $U^+_{\bar{a}_0}$ of $\bar{a}_0$ where $\frac{d}{dt}I(\bar{a},t)$ is positive in  of $U^+_{\bar{a}_0}\cap E_0^C$.

\section{Calculating partial derivatives}\label{Taylor}

A direct calculation of the derivatives of $\frac{d}{dt}I(\bar{a},t)$ with respect to the $a_i$ can be demanding since the eigenvalues of $\rho$ are the roots of a degree $\dim (\mathcal{H}_A\otimes\mathcal{H}_S)$ polynomial.
To avoid this difficulty we use a method for calculating the derivatives and second derivatives in a point $\bar{a}$ adapted from Ref. \cite{Tsing}. The method given there is valid for real symmetric matrices but the generalization to Hermitian complex matrices is straightforward. 
We describe this method in the following paragraphs.

Let $f$ be a spectral function defined on a set of $n\times n$ Hermitian matrices $A$ parametrized by real numbers $a_i$. By spectral function we mean a function that only depends on the eigenvalues $\{\lambda_k\}_{k=1}^n$ of $A$ but not on the ordering of the eigenvalues. Furthermore, assume that $f$ is analytic in the point $\bar{a}$ and let $u_k(\bar{a})$ be the normalized eigenvector of $A(\bar{a})$ corresponding to the eigenvalue $\lambda_k(\bar{a})$. 

Then the first and second order partial derivatives of $f$ with respect to the parameters $a_i$ in point $\bar{a}$ can be expressed as

\begin{eqnarray}
\frac{\partial f(\bar{a})}{\partial a_i}=\sum_{k}\frac{\partial{f[\lambda(\bar{a})]}}{\partial \lambda_k}h_i^k(\bar{a}),
\end{eqnarray}
and
\begin{eqnarray}
\frac{\partial^2 f(\bar{a})}{\partial a_i \partial a_j}=&&\sum_{k,l}\frac{\partial^2{f[\lambda(\bar{a})]}}{\partial \lambda_k \partial \lambda_l}h_i^k(\bar{a})h_j^l(\bar{a})\nonumber\\&&+\sum_{k}\frac{\partial{f[\lambda(\bar{a})]}}{\partial \lambda_k}h_{ij}^k(\bar{a})+\eta_{ij}(\bar{a}),
\end{eqnarray}
respectively, where
\begin{eqnarray}
h_i^k(\bar{a})=&&u_k^\dagger\frac{\partial A(\bar{a})}{\partial a_i}u_k,\nonumber\\
h_{ij}^k(\bar{a})=&&u_k^\dagger\frac{\partial^2 A(\bar{a})}{\partial a_i \partial a_j}u_k+\sum_{l|\lambda_k\neq \lambda_l}\frac{\alpha_{ij}^{kl}(\bar{a})}{\lambda_k(\bar{a})-\lambda_l(\bar{a})},\nonumber\\
\alpha_{ij}^{kl}(\bar{a})=&&\left(u_k^\dagger(\bar{a})\frac{\partial A(\bar{a})}{\partial a_i}u_l(\bar{a})\right)\left(u_l^\dagger(\bar{a})\frac{\partial A(\bar{a})}{\partial a_j}u_k(\bar{a})\right)\nonumber\\&&+\left(u_k^\dagger(\bar{a})\frac{\partial A(\bar{a})}{\partial a_j}u_l(\bar{a})\right)\left(u_l^\dagger(\bar{a})\frac{\partial A(\bar{a})}{\partial a_i}u_k(\bar{a})\right),\nonumber\\
\eta_{ij}(\bar{a})=&&\sum_{k,l|\lambda_k=\lambda_l,k<l}\alpha_{ij}^{kl}(\bar{a})\frac{\partial^2 f[\lambda(\bar{a})]}{\partial^2 \lambda_k}.
\end{eqnarray}
Note that when some eigenvalues coincide the choice of eigenvectors is not unique. However, while e.g. $h_i^k$ depends on this choice the partial derivatives themselves are independent and can be evaluated using any choice of eigenvectors.

When the diagonal form of $A$ and the eigenvectors $u_k(\bar{a})$ are known the method described here can greatly simplify the computation of the partial derivatives.

\section{Mutual information for random unitary dynamics}\label{mutual}
We here show that the mutual information is non-increasing for some cases of non CP-divisible random unitary qubit dynamics by studying a neighbourhood of the stationary states using the methods described in Appendix \ref{meth} and Appendix \ref{Taylor}. 

Random unitary dynamics for a qubit is defined by the dynamical maps

\begin{eqnarray}
\Lambda_t(\sigma_x)&&=e^{-\int_{0}^{t}(\gamma_z(\tau)+\gamma_y(\tau))d\tau}\sigma_x,\nonumber\\
\Lambda_t(\sigma_y)&&=e^{-\int_{0}^{t}( \gamma_z(\tau)+\gamma_x(\tau))d\tau}\sigma_y,\nonumber\\
\Lambda_t(\sigma_z)&&=e^{-\int_{0}^{t}(\gamma_x(\tau)+\gamma_y(\tau))d\tau}\sigma_z,\nonumber\\
\Lambda_t(\mathbbm{1})&&=\mathbbm{1},
\end{eqnarray}
where $\gamma_k(t)$ are real valued functions of $t$. The dynamical maps are bijective for all $t$ and the intermediate maps are given by
\begin{eqnarray}\label{ranu}
V_{s,t}(\sigma_x)&&=e^{-\int_{t}^{s}(\gamma_z(\tau)+\gamma_y(\tau))d\tau}\sigma_x,\nonumber\\
V_{s,t}(\sigma_y)&&=e^{-\int_{t}^{s}(\gamma_z(\tau)+\gamma_x(\tau))d\tau}\sigma_y,\nonumber\\
V_{s,t}(\sigma_z)&&=e^{-\int_{t}^{s}(\gamma_x(\tau)+\gamma_y(\tau))d\tau}\sigma_z,\nonumber\\
V_{s,t}(\mathbbm{1})&&=\mathbbm{1}.
\end{eqnarray}
The corresponding generator of the dynamics is
\begin{eqnarray}
\mathcal{L}_t(\rho)=\sum_{k=x,y,z}\gamma_k(t)(\sigma_k\rho\sigma_k-\rho).
\end{eqnarray}
The dynamics is CP-divisible if and only if $\gamma_k(t)\geq 0$ for all $k$. Moreover, the dynamics is P-divisible if and only if the conditions

\begin{eqnarray}\label{rattes}
\gamma_y(t)+\gamma_z(t)\geq{0},\nonumber\\
\gamma_x(t)+\gamma_z(t)\geq{0},\nonumber\\ 
\gamma_x(t)+\gamma_y(t)\geq{0},
\end{eqnarray}
are satisfied since the intermediate maps are then contractive in the trace norm \cite{koss1,koss2,ruskai}. 

We consider an ancilla that is also a qubit and explicitly introduce coordinates $a_i$ for $B(\mathcal{H}_A\otimes \mathcal{H}_S)$ with respect to an orthonormal basis $\{e_i\}_{i=0}^{15}$ defined by

\begin{align}
&e_0=\mathbbm{1}\otimes{\mathbbm{1}},& &e_{8}=\sigma_y\otimes{\mathbbm{1}},\nonumber\\
&e_{1}=\mathbbm{1}\otimes{\sigma_x},& &e_9=\sigma_y\otimes{\sigma_x},\nonumber\\
&e_{2}=\mathbbm{1}\otimes{\sigma_y},& &e_{10}=\sigma_y\otimes{\sigma_y},\nonumber\\
&e_{3}=\mathbbm{1}\otimes{\sigma_z},& &e_{11}=\sigma_y\otimes{\sigma_z},\nonumber\\
&e_4=\sigma_x\otimes{\mathbbm{1}},& &e_{12}=\sigma_z\otimes{\mathbbm{1}},\nonumber\\
&e_5=\sigma_x\otimes{\sigma_x},& &e_{13}=\sigma_z\otimes{\sigma_x},\nonumber\\
&e_6=\sigma_x\otimes{\sigma_y},& &e_{14}=\sigma_z\otimes{\sigma_y},\nonumber\\
&e_7=\sigma_x\otimes{\sigma_z},& &e_{15}=\sigma_z\otimes{\sigma_z},
\end{align}
where all operators are of the form $\chi_A\otimes{\chi_S}$ for $\chi_A\in B(\mathcal{H}_A)$  and $\chi_S \in B(\mathcal{H}_S)$.
A state $\rho$ is represented as

\begin{eqnarray}
\rho=\frac{1}{4} \mathbbm{1}\otimes{\mathbbm{1}}+\sum_{i=1}^{15} a_i e_i,
\end{eqnarray}
where $a_i=\frac{1}{4}\Tr(\rho e_i)$.

We begin the analysis of $\frac{d}{dt}I(\bar{a},t)$ in the neighbourhood of the stationary states by considering $\Int[S(\mathcal{H}_A\otimes\mathcal{H}_S)]$ where $\frac{d}{dt}I(\bar{a},t)$ is analytic.
We calculate the first and second derivatives of $\frac{d}{dt}I(\bar{a},t)$ at the stationary states in the interior of the set of states and find the eigenvalues of the Hessian matrix. On the subset of states that fall in the zero eigenspace of the Hessian we then directly evaluate  $\frac{d}{dt}I(\bar{a},t)$. Finally, we describe the neighbourhood of the intersection of the stationary states with the boundary of the set of states.

The stationary states are of the form $1/2\rho_A\otimes{\mathbbm{1}}$ for arbitrary $\rho_A$. 
For these states all first derivatives $\frac{d}{dt}I(\bar{a},t)$ with respect to $\bar{a}$ are zero.  Therefore, there exists some sufficiently small neighbourhood of the set of stationary states where the second order terms of the Taylor expansion in $\bar{a}$ determines the sign of $\frac{d}{dt}I(\bar{a},t)$, in every direction where the second derivative is non-zero.
For the purpose of calculating these derivatives we note that unitary transformations on the ancilla do not change the mutual information and it is sufficient to consider diagonal $\rho_A$. Thus, the purity of the state of the ancilla is the only relevant parameter.
The diagonal stationary states are of the form $\frac{1}{4} \mathbbm{1}\otimes{\mathbbm{1}}+a_{12}\sigma_z\otimes{\mathbbm{1}}$ for $-1/4\leq a_{12}\leq{1/4}$. The states for which  $-1/4 < a_{12} < {1/4}$ are in $\Int [S(\mathcal{H}_A\otimes \mathcal{H}_S)]$ and the states with coordinates $a_{12}=\pm 1/4$ are at the boundary of the set of states.

The second derivatives at the diagonal stationary states were calculated using the method described in Appendix \ref{Taylor} and the Hessian matrix was diagonalized. The Hessian has 6 eigenvalues that are identically zero for all stationary states in  $\Int [S(\mathcal{H}_A\otimes \mathcal{H}_S)]$ regardless of the values of the parameters $\gamma_k(t)$ and 9  eigenvalues that can take non-zero values.
These 9 eigenvalues are 

\begin{eqnarray}\label{eig}
32[\gamma_y(t) + \gamma_z(t)]\left (\frac{16 a_{12}^2+1}{ 16 a_{12}^2-1 }\right),\nonumber\\ 
 32[\gamma_x(t) + \gamma_z(t)]\left (\frac{16 a_{12}^2+1}{ 16 a_{12}^2-1 }\right),\nonumber\\ 
 32[\gamma_x(t) + \gamma_y(t)]\left (\frac{16 a_{12}^2+1}{ 16 a_{12}^2-1 }\right),\nonumber\\
 -8 [\gamma_y(t) + \gamma_z(t)]\frac{\atanh(4 a_{12})}{a_{12}},\nonumber\\ 
 -8 [\gamma_y(t) + \gamma_z(t)]\frac{\atanh(4 a_{12})}{a_{12}},\nonumber\\ 
 -8 [\gamma_x(t) + \gamma_z(t)]\frac{\atanh(4 a_{12})}{a_{12}},\nonumber\\ 
-8 [\gamma_x(t) + \gamma_z(t)]\frac{\atanh(4 a_{12})}{a_{12}},\nonumber\\ 
 -8 [\gamma_x(t) + \gamma_y(t)]\frac{\atanh(4 a_{12})}{a_{12}},\nonumber\\ 
 -8 [\gamma_x(t) + \gamma_y(t)]\frac{\atanh(4 a_{12})}{a_{12}}.
\end{eqnarray}
The eigenvalues in Eq. (\ref{eig}) are all non-positive if and only if the conditions in Eq. (\ref{rattes}) are satisfied, i.e., if and only if the dynamics is P-divisible. In particular they are all strictly negative if $\gamma_i(t)+\gamma_j(t)>{0}$ for all $i,j$.
%If any of the three inequalities in Eq. (\ref{rattes}) is not satisfied the three eigenvalues containing the corresponding factor \red are \black positive. 
%We can conclude that if the conditions in Eq. (\ref{rattes}) are satisfied
In this case there exists a sufficiently small neighbourhood of the stationary states where $\frac{d}{dt}I(\bar{a},t)$ is negative in all the directions that have a non-zero component orthogonal to the zero-eigenspace of the Hessian .

Next, we investigate $\frac{d}{dt}I(\bar{a},t)$ on the eigenspace of the eigenvalues that are identically zero, where its sign is determined by higher order derivatives. Here it is straightforward to evaluate $\frac{d}{dt}I(\bar{a},t)$ directly. 
The zero eigenspace $E_0(a_{12})$ as a function of $a_{12}$, is spanned by the six vectors
$(\mathbbm{1}+4a_{12}\sigma_z)\otimes{\sigma_i}$ and $\sigma_i\otimes{\mathbbm{1}}$ for $i=x,y,z$. These vectors are tangent to the set of product states, but the tangent plane $E_0(a_{12})$ also contains correlated states. 
Consider the point $\frac{1}{4} \mathbbm{1}\otimes{\mathbbm{1}}+a_0\sigma_z\otimes{\mathbbm{1}}$ on the set of stationary states. The states in the subspace $E_0(a_0)$ are of the form

\begin{eqnarray}
\frac{1}{4} \mathbbm{1}\otimes{\mathbbm{1}}+(\mathbbm{1}+4 a_{0}\sigma_z)\otimes{(a_{1}\sigma_x+a_{2}\sigma_y+a_{3}\sigma_z)}\nonumber\\
+ (a_4\sigma_x+ a_{8}\sigma_y+a_{12}\sigma_z)\otimes{\mathbbm{1}}.
\end{eqnarray}
Since the mutual information is independent of unitary transformations on the system we can diagonalize $a_{1}\sigma_x+a_{2}\sigma_y+a_{3}\sigma_z$. Let $\pm\lambda(s)=\pm\sqrt{a_{1}^2(s)+a_{2}^2(s)+a_{3}^2(s)}$ be the corresponding eigenvalues as functions of time where

\begin{eqnarray}
a_1(s)&&=a_{1}e^{-\int_{t}^{s}(\gamma_z(\tau)+\gamma_y(\tau))d\tau},\nonumber\\
a_2(s)&&=a_{2}e^{-\int_{t}^{s}(\gamma_z(\tau)+\gamma_x(\tau))d\tau},\nonumber\\
a_3(s)&&=a_{3}e^{-\int_{t}^{s}(\gamma_x(\tau)+\gamma_y(\tau))d\tau}.
\end{eqnarray}
The density matrix is now block-diagonal and the characteristic polynomial factorizes into two quadratic polynomials. The mutual information $I[E_0(a_0)]$ calculated from the corresponding eigenvalues, as a function on $E_0(a_0)$, is  
\begin{eqnarray}
I[E_0(a_0)]=&&\left(\frac{1}{4}-\lambda(s)-\omega_-\right) \ln\left(\frac{1}{4} - \lambda(s) - \omega _-\right)\nonumber\\
      &&+ \left(\frac{1}{4}-\lambda(s)+\omega_- \right)\ln\left(
   \frac{1}{4} - \lambda(s) + \omega_-\right)\nonumber\\
      &&+ \left(\frac{1}{4} + \lambda(s) + \omega_+ \right)\ln\left(
   \frac{1}{4} + \lambda(s) + \omega_+\right)\nonumber\\
       &&+ \left(\frac{1}{4} + \lambda(s) - \omega_+ \right)\ln\left(
   \frac{1}{4} + \lambda(s) - \omega_+\right)\nonumber\\ &&- \left(\frac{1}{2} + 
    2 \lambda(s)\right)\ln\left(\frac{1}{2} + 2 \lambda(s)\right)\nonumber\\ && - \left(\frac{1}{2} - 2 \lambda(s)\right) \ln\left(\frac{1}{2} - 2 \lambda(s)\right)\nonumber\\ &&- \left(\frac{1}{2} + 
    2 \eta\right)\ln\left(\frac{1}{2} + 2 \eta\right)\nonumber\\ && - \left(\frac{1}{2} - 2 \eta\right) \ln\left(\frac{1}{2} - 2 \eta\right),
\end{eqnarray}
where $\omega_{\pm}=\sqrt{a_{4}^2 + a_{8}^2+[a_{12} \pm 4 a_0 \lambda(t)]^2}$ and $\eta=\sqrt{a_{4}^2 + a_{8}^2+a_{12}^2}$.
Since the only dependence of $s$ in $I[E_0(a_0)]$ is in $\lambda(s)$, the time derivative of the mutual information can be expressed as $\frac{d I[E_0(a_0)]}{dt}=\frac{d I[E_0(a_0)]}{d\lambda(s)}\frac{d \lambda(s)}{ds}\big{|}_{s=t}$, where $\frac{d \lambda(s)}{ds}\big{|}_{s=t}$ has the form

\begin{eqnarray}\label{kk}
\frac{a_{1}^2[\gamma_z(t)+\gamma_y(t)]+a_{2}^2[\gamma_x(t)
+\gamma_z(t)]+a_{3}^2[\gamma_x(t)+\gamma_y(t)]
}{\sqrt{a_{1}^2+a_{2}^2+a_{3}^2}}.
\nonumber\\
\end{eqnarray}
When the conditions in Eq. (\ref{rattes}) are satisfied, i.e., when the dynamics is P-divisible, $\frac{d \lambda(s)}{ds}\big{|}_{s=t}$ is non-negative for all $a_{1},a_{2},a_{3}$.
Since $\frac{d I[E_0(a_0)]}{dt}\leq 0$ for all $\bar{a}\in E_0(a_0)$ when the dynamics is CP-divisible it follows that $\frac{d I[E_0(a_0)]}{d\lambda(s)}\big{|}_{s=t}$ is non-positive for all $\bar{a}\in E_0(a_0)$. Therefore we can conclude that $\frac{d I[E_0(a_0)]}{dt}\leq 0$ for all $\bar{a}\in E_0(a_0)$ when $V_{s,t}$ is P-divisible.

The above analysis shows that there exist non-Markovian P-divisible dynamics for which there is a neighbourhood of the stationary states in $\Int[S(\mathcal{H}_A\otimes\mathcal{H}_S)]$ where no increase of the mutual information occurs. 
It remains to consider the neighbourhood of the set of stationary states in the boundary of the set of states, i.e., the neighbourhood of $1/4(\mathbbm{1}+\sigma_z)\otimes{\mathbbm{1}}$ and $1/4( \mathbbm{1}-\sigma_z)\otimes{\mathbbm{1}}$.
The states for which $a_{12}=\pm 1/4$ are of the form $1/4( \mathbbm{1}\pm \sigma_z)\otimes\rho$, where $\rho\in B(\mathcal{H}_S)$.
This can be seen by noting that 
if $a_{12}=\pm 1/4$, it follows that $a_4=a_8=0$ to ensure non-negative eigenvalues of the reduced state on $\mathcal{H}_A$. Thus, for such states the reduced state of the ancilla is pure, which implies that all states in this neighbourhood of $1/4( \mathbbm{1}\pm \sigma_z)\otimes\rho$ are product states.
Since any product state has zero mutual information and remains a product state during the evolution it follows that $\frac{d}{dt}I(\bar{a},t)$ is zero for all states in any neighbourhood of $1/4( \mathbbm{1}\pm \sigma_z)\otimes{\mathbbm{1}}$ where $a_{12}=\pm 1/4$.

Finally, we can conclude that there exist non-Markovian P-divisible dynamics for which there is a neighbourhood of the stationary states where no increase in the mutual information occurs. Moreover, the rates $\gamma_k(\tau)$ for $0\leq\tau\leq t$ can be chosen such that the image of $\Lambda_t$ is contained in this neighbourhood. Therefore, there exist evolutions for which the non-Markovian character can not be witnessed by an increase in the mutual information.

\section{Tuning the rates to resize the image of $\Lambda_t$}\label{tuning}
Here we describe how the image of $\Lambda_t$ for a random unitary dynamics can always be contained in a given neighbourhood of the stationary states by tuning the rates $\gamma_k$.

Consider the random unitary dynamics defined by

\begin{eqnarray}\label{ranu}
\Lambda_{t}(\sigma_x)&&=e^{-\int_{0}^{t}(\gamma_z(\tau)+\gamma_y(\tau))d\tau}\sigma_x,\phantom{u}
\Lambda_{t}(\sigma_y)=e^{-\int_{0}^{t}(\gamma_z(\tau)+\gamma_x(\tau) )d\tau}\sigma_y,\nonumber\\
\Lambda_{t}(\sigma_z)&&=e^{-\int_{0}^{t}(\gamma_x(\tau)+\gamma_y(\tau))d\tau}\sigma_z,\phantom{u}
\Lambda_{t}(\mathbbm{1})=\mathbbm{1}.
\end{eqnarray}
For any $\epsilon>0$ we can choose the functions $\gamma_k(\tau)$ such that $e^{-\int_{0}^{t}(\gamma_i(\tau)+\gamma_j(\tau))d\tau}<\epsilon$ for all $i\neq j$. Moreover, the value of the integral $\int_{0}^{t}(\gamma_i(\tau)+\gamma_j(\tau))d\tau$ can be made arbitrarily large independently of the $\gamma_k(t)$. This can be done for example by choosing $\gamma_k(\tau)$ such that the integral $\int_{t_1}^{t_2}\gamma_k(\tau)d\tau>-\ln (\epsilon)$  where $0<t_1<t_2<t$, for each $k$.

Therefore, for any neighbourhood of the stationary states at time $t$ and any $\gamma_k(t)$ we can choose the $\gamma_k(\tau)$ for $0<\tau<t$ such that the image of $\Lambda_t$ is contained in this neighbourhood.

%and let $\rho_A\otimes \mathbbm{1}+\sum_k c_k\chi_k\otimes \sigma_k$ be an arbitrary initial state.

%At time $t$ the state is 

%{eqnarray}\rho_A\otimes\mathbbm{1}+ c_xe^{-\int_{0}^{t}(\gamma_z(\tau)+\gamma_y(\tau))d\tau}\chi_x\otimes\sigma_x+c_ze^{-\int_{0}^{t}(\gamma_x(\tau)+\gamma_y(\tau))d\tau}\chi_z\otimes\sigma_z+c_ye^{-\int_{0}^{t}(\gamma_z(\tau)+\gamma_x(\tau))d\tau}\chi_y\otimes\sigma_y.
%%For any $c>0$ we can choose the functions $\gamma_k(t)$ such that

%Now consider a function $g(t)=f(t)+1$ where $f(t)$ is non-zero on the interval $t_1,t_2$, for $0<t_1<t_2<t$, and zero elsewhere. By tuning the rates $\gamma_k(t)\arrow \gamma_k(t)$

%\section{$\Pi_A(\rho_{AB}) \neq \{ \mathbf{0}\}$ }\label{neverempty}
\section{The set of maximally entropic measurements is non-empty}\label{neverempty}
We  explicitly construct an element $\{ P_{i}\}_i$ of $\Pi(\rho)$ for an arbitrary state $\rho$. The method that we use should convince the reader that there are innumerable other ways to construct a ME-POVM with any number of outputs.

By definition $\{ P_{i}\}_{i=1,\dots,n} \in\Pi(\rho)$ if the output ensemble $\mathcal{E}(\rho, \{ P_{i}\}_i)=\{p_i, \rho_{i} \}_{i}$ is characterized by $p_i=1/n$. In general, we have that $p_{i} = \tr{ \rho P_{i} } \, ,$. Using an orthogonal decomposition of $\rho$, we can always write it as $\rho  = \sum_{i=1}^{d} \pi_i \ketbra{i}{i}$, where $\{\ket{i}\}_i$ is an orthonormal basis of the Hilbert space $\mathcal{H}$. The condition $\sum_{i=1}^{d} \pi_i =1$ implies that there exist an $\overline{i}$, such that $S (\overline{i}) \equiv \sum_{i=1}^{\overline{i}} \pi_i > 1/2$ and $S(\overline{i}-1) \equiv \sum_{i=1}^{\overline{i}-1} \pi_i \leq 1/2$. We consider the following class of 2-output POVM that depends on a real parameter $\omega \in [0,1]$:
$
P_{1} (\omega) = \sum_{i=1}^{\overline{i}-1} \ketbra{i}{i} + \omega \ketbra{\overline{i}}{\overline{i}} \, ,
$
$
P_{2}(\omega) = (1-\omega)\ketbra{\overline{i}}{\overline{i}}+ \sum_{i= \overline{i}+1}^{d}\ketbra{i}{i} \, .
$
We evaluate $p_{1}$ for a general value of $\omega$ and we obtain:
$
p_{1}(\omega) =\sum_{i=1}^{\overline{i}-1} \pi_i + \omega \, \pi_{\overline{i} } = S(\overline{i}-1) + \omega \, \pi_{\overline{i} }  \, .
$
It is clear that, since $p_{1}(0) = S(\overline{i}-1)\leq1/2$ and $p_{1}(1) = S(\overline{i})>1/2$, the value $\omega =  \overline{\omega}\equiv (1/2 - S(\overline{i}-1) )/{\pi_{\overline{i}} }$, gives the uniform distribution  $p_{1,2}(\overline{\omega})=1/2$ and consequently $\{ P_{i}(\overline \omega)  \}_i \in \Pi(\rho)$, i.e,, is a  ME-POVM for $\rho$.

\section{Monotonic behaviour of $C$  {and $C^{(n)}$} under local operations}\label{monolocal}

Firstly, we prove that {$C_A$} is monotone under local operations of the form $\Lambda_A \otimes  \mathcal{I}_B$, and secondly we consider the case where the local operation is $\mathcal{I}_A \otimes  \Lambda_B$, where $\Lambda_A$ ($\Lambda_B$) is a CPTP map on $A$ ($B$) and $\mathcal{I}_A$ ($\mathcal{I}_B$) is the identity map on $A$ ($B$). The proof for {$C_A$} easily generalizes to {$C_B$} and {$C$}.
{Finally, we prove that the same monotonicity property holds for $C^{(n)}$ for any $n\geq 2$. We denote the set of ME-POVMs acting on $A$ for the state $\rho_{AB}$ by $\Pi_A(\rho_{AB})$ and similarly for $B$.}

In order to show the effect  of the application of a local operation of the form  $\Lambda_A \otimes  \mathcal{I}_B$ on $C_A(\rho_{AB})$, we look at $\Pi_A(\rho_{AB})$ in a different way. Each element of this collection is a ME-POVM for $\rho_{AB}$, i.e. they generate sets of { \it equiprobable ensembles of states} (EES) from $\rho_{AB}$. In fact 
\begin{equation}\label{CAapp}
C_A (\rho_{AB}) \equiv \max_{ \left\{ P_{A,i}\right\}_i  \in \Pi_A \left( \rho_{A} \right) }  P_g \left( \mathcal{E} \left( \rho_{AB} ,\left\{ P_{A,i} \right\}_i \right)  \right)  - \frac{1}{2} \, .
\end{equation}
 is a maximization over all the possible EES that we can generate from $\rho_{AB}$ with a measurement procedure on $A$.

The effect of the first local operation that we consider is: $\tilde{\rho}_{AB}=  \Lambda_A \otimes  \mathcal{I}_B \, ( \rho_{AB} ) = \sum_k \left( E_k \otimes \mathbbm{1}_B \right) \cdot \rho_{AB} \cdot \left( E_k \otimes \mathbbm{1}_B \right)^\dagger  \, ,$
where $\left\{ E_k \right\}_k$ is the set of the Kraus operators that defines $\Lambda_A$. What is the relation between $\Pi_A (\rho_{AB}) $ and $\Pi_A (\tilde{\rho}_{AB})$? Given an $n$-output ME-POVM for $\tilde \rho_{AB}$, i.e. $\{ P_{A,i}\}_i \in \Pi_A (\tilde{\rho}_{AB})$, the probabilities and the states of the output ensemble $\mathcal{E} \left( \tilde{\rho}_{AB}, \{ P_{A,i} \}_i\right)$ are $\tilde{p}_i =  \tr{ \tilde{\rho}_{AB} \cdot P_{A,i} }=1/n$ and $\tilde \rho_{B,i} = \trA{\tilde\rho_{AB} \cdot P_{A,i}}/\tilde{p}_i$. Now we look at the term
$$
\trA{\tilde{\rho}_{AB} \cdot P_{A,i}} = \tr{\Lambda_{A}\otimes \mathcal{I}_B\, ( \rho_{AB}) \cdot P_{A,i}} =
$$
$$
=\trA{\sum_k (E_k \otimes \mathbbm{1}_B)\cdot \rho_{AB} \cdot (E_k^\dagger \otimes \mathbbm{1}_B)\cdot P_{A,i} }=
$$
$$
=\trA{\rho_{AB} \sum_k (E_k^\dagger \otimes \mathbbm{1}_B) \cdot P_{A,i} \cdot (E_k \otimes \mathbbm{1}_B) } =
$$
$$
= \trA{\rho_{AB} \cdot \Lambda^*_{A} (P_{A,i} ) } = \trA{\rho_{AB} \cdot \tilde P_{A,i} } \, ,
$$
and we rewrite the probabilities and the output states as:
$\tilde p_i= \mbox{Tr}[ \rho_{AB} \cdot \tilde P_{A,i} ] = 1/n$ and $\rho_{B,i}=\mbox{Tr}_A[ \rho_{AB} \cdot \tilde P_{A,i} ] /\tilde p_i$. This ensemble is an EES. Next we show that:
$\{ \tilde{P}_{A,i} \}_i =\left\{ \Lambda^*_A \left( P_{A,i} \right) \right\}_i = \{ \sum_k E_k^\dagger\cdot P_{A,i} \cdot E_k \}_i \, ,
$
is a POVM. The elements of $\{ \tilde{P}_{A,i} \}_i$ sum up to the identity: 
$
\sum_i \tilde{P}_{A,i} =   \sum_{k,i} E_k^\dagger \, P_{A,i} \,  E_k = \sum_{k} E_k^\dagger \, \left( \sum_i P_{A,i} \right) \, E_k = \sum_{k} E_k^\dagger\,  E_k = \mathbbm{1}_B \, ,
$
and they are positive operators:
$
\tilde{P}_{A,i} = \sum_k E_k^\dagger\,  P_{A,i} \,  E_k = \sum_{k}  E_k^\dagger \,  M^\dagger_{A,i} \, M_{A,i} \,  E_k =  \tilde{M}_{A,i}^\dagger \tilde{M}_{A,i} \, ,
$
where the decomposition $P_{A,i} = M^\dagger_{A,i} M_{A,i}$ exists since $P_{A,i}$ is positive-semidefinite and  $\tilde{M}_{A,i} = \sum_k M_{A,i} \, E_k$. It follows that, $\{ \tilde{P}_{A,i} \}_i$ is a ME-POVM for $\rho_{AB}$, i.e. $\{ \tilde{P}_{A,i} \}_i\in \Pi_A(\rho_{AB})$.  Thus, for every ME-POVM $\{P_{A,i}\}_i \in \Pi_A(\tilde\rho_{AB})$ for $\tilde \rho_{AB}$, there is a ME-POVM $\{\tilde P_{A,i}\}_i \in \Pi_A(\rho_{AB})$ for $\rho_{AB}$, such that the output ensembles are identical: $\mathcal{E} (\tilde \rho_{AB}, \{P_{A,i}\}_i) = \mathcal{E} (\rho_{AB}, \{\tilde P_{A,i} \}_i )$.
Thus, any EES that can be generated from $\tilde{\rho}_{AB}$, is obtainable from $\rho_{AB}$ as well
\begin{equation}\label{EI}
\bigcup_{ \{P_{A,i} \}_i \in \Pi_A (\tilde{\rho}_{AB} ) }   \!\!\!\!\!\!\!\!  \mathcal{E} \left( \tilde{\rho}_{AB}, \, \{ P_{A,i} \}_i \right)\,  \subseteq \!\!\!\!\!\!\! \bigcup_{ \{P_{A,i} \}_i \in \Pi_A ({\rho}_{AB} ) }    \!\!\!\!\!\!\!\! \mathcal{E} \left( {\rho}_{AB}, \, \{ P_{A,i} \}_i \right) \, .
\end{equation}
Finally, because $C_A(\rho_{AB})$ could be thought as the maximum guessing probability of the EESs that can be generated from $\rho_{AB}$ (see Eq. (\ref{CAapp})), we conclude that 
\begin{equation}\label{monA2}
C_A \left( \rho_{AB} \right) \geq C_A \left(  \Lambda_A \otimes  \mathcal{I}_B \,  ( \rho_{AB}) \right) \, ,
\end{equation}
for any state $\rho_{AB}$ and CPTP map $\Lambda_A$.

{Fixing the number $n$ of outputs of the ME-POVMs considered in (\ref{CAapp}), Eq. (\ref{EI}) becomes:
\begin{equation}\label{EIn}
\bigcup_{ \{P_{A,i} \}_{i=1}^n \in \Pi_A (\tilde{\rho}_{AB} ) }   \!\!\!\!\!\!\!\!  \mathcal{E} \left( \tilde{\rho}_{AB}, \, \{ P_{A,i} \}_i \right)\,  \subseteq \!\!\!\!\!\!\! \bigcup_{  \{P_{A,i} \}_{i=1}^n \in \Pi_A ({\rho}_{AB} ) }    \!\!\!\!\!\!\!\! \mathcal{E} \left( {\rho}_{AB}, \, \{ P_{A,i} \}_i \right) \, .
\end{equation}
Therefore, it follows that:
\begin{equation}\label{monA2n}
C_A^{(n)} \left( \rho_{AB} \right) \geq C_A^{(n)} \left(  \Lambda_A \otimes  \mathcal{I}_B \,  ( \rho_{AB}) \right) \, ,
\end{equation}
for any integer $n\geq 2$, state $\rho_{AB}$ and  CPTP map $\Lambda_A$.
}

Next we show the property of monotonicity of $C_A(\rho_{AB})$  under  the action of local operations of the form $\mathcal{I}_A \otimes  \Lambda_B$. We find that the collection of the ME-POVMs for $\tilde{\rho}_{AB} =  \mathcal{I}_A \otimes  \Lambda_B \, ( {\rho}_{AB})$, i.e. $\Pi_A (\tilde{\rho}_{AB})$, coincides with $\Pi_A (\rho_{AB})$. 

In order to prove this, we apply a general POVM $\{P_{A,i}\}_i$ on both $\rho_{AB}$ and $\tilde \rho_{AB}$ and we show that the respective output ensembles are defined by the same probability distribution. We can write $p_i= \tr{\rho_{AB}\cdot P_{A,i}}$ and $\tilde p_i =\tr{ \mathcal{I}_A\otimes \Lambda_B \, ( \rho_{AB}) \cdot P_{A,i} } = \tr{\rho_{AB} \cdot P_{A,i} }$, where the last step uses the trace-preserving property of the superoperator $\mathcal{I}_A\otimes \Lambda_B$. Consequently, $p_i=1/n$ if and only if $\tilde p_i=1/n$ and $\{ P_{A,i}\}_i\in \Pi_A(\rho_{AB})$ if and only if $\{ P_{A,i}\}_i\in \Pi_A(\tilde \rho_{AB})$
\begin{equation}\label{uguale}
\Pi_A (\rho_{AB} ) = \Pi_A (\tilde{\rho}_{AB}) \, .
\end{equation}
Given a ME-POVM for both $\rho_{AB}$ and $\tilde{\rho}_{AB}$, we relate the output states
\begin{equation}\label{rhobicontratti}
\tilde{\rho}_{B,i} = \Lambda_B \cdot \trA{ \rho_{AB} P_{A,i}  }  / p_i = \Lambda_B (\rho_{B,i}) \, .
\end{equation}
From Eq. (\ref{rhobicontratti}) and the definition of the guessing probability, it follows that
\begin{equation}\label{monB0}
P_g\left(  \left\{ p_i, \, \rho_{B,i}  \right\}_i  \right) \geq P_g\left(  \left\{ p_i, \, \Lambda_B (\rho_{B,i})  \right\}_i  \right) \, ,
\end{equation}
and, considering Eq. (\ref{uguale}), Eq. (\ref{rhobicontratti}) and Eq. (\ref{monB0})
\begin{equation}\label{monB}
C_A \left( \rho_{AB} \right) \geq C_A \left(  \mathcal{I}_A \otimes \Lambda_B   \,  ( \rho_{AB}) \right) \, ,
\end{equation}
that is true for any state $\rho_{AB}$ and CPTP map $\Lambda_B$.

{From Eq. (\ref{uguale}) it follows the collection of the $n$-output ME-POVMs does not change if we apply a CPTP map $\Lambda_B$ on $\rho_{AB}$. Therefore, since Eq. (\ref{monB0}) is true for any number of outputs:
\begin{equation}\label{monB}
C_A^{(n)} \left( \rho_{AB} \right) \geq C_A^{(n)}  \left(  \mathcal{I}_A \otimes \Lambda_B   \,  ( \rho_{AB}) \right) \, ,
\end{equation}
for any integer $n\geq 2$, state $\rho_{AB}$ and  CPTP map $\Lambda_B$.
}

We underline that from this proof we automatically obtain the invariance under local unitary transformations of $C$ and $C^{(n)}$ for any $n\geq 2$.

\section{Proof that $C_A({\rho}_{AB}^{(\tau)}) \geq  C^{(2)}_B({\rho}_{AB}^{(\tau)})$ }\label{CB2CA}

In this appendix (where {from now on} we omit the time dependence of $\rho_{AB}^{(\tau)}(t)$, $\rho_B'^{(\tau)}(t)$ and $\rho_B''^{(\tau)}(t)$) we show that $ C_A( \rho_{AB}^{(\tau)}) \geq C^{(2)}_B( \rho_{AB}^{(\tau)})$, where $C^{(2)}_B(\rho_{AB}^{(\tau)})$ is defined by
$$
C_B^{(2)} ({\rho}_{AB}^{(\tau)}) =\!\!\!\!\!\!\!\! \max_{ \left\{P_{B,i}\right\}_i \in \Pi_B^{(2)} \left(  \rho_{AB}^{(\tau)} \right) } \!\!\!\!P_g \left( \mathcal{E} \left( \rho_{AB}^{(\tau)} ,\left\{ P_{B,i} \right\}_i \right)  \right)-\frac{1}{2} \, ,
$$
where $\Pi_B^{(2)} (  \rho_{AB}^{(\tau)} )$ is the set of the 2-output ME-POVMs acting on $B$.
 In Appendix  \ref{2enough} we show that $C^{(2)}_B( \rho_{AB}^{(\tau)}) = C_B( \rho_{AB}^{(\tau)})$ and this completes the proof that $C_A({\rho}_{AB}^{(\tau)}) \geq  C_B({\rho}_{AB}^{(\tau)})$.

We  apply a general but fixed 2-output ME-POVM for ${\rho}_{AB}^{(\tau)}$, where now the measured system is  $B$: $\{P_{B,i}^{(2)}\}_i = \{ P_{B}, \, \overline{P}_B \} \in \Pi_B({\rho}_{AB}^{(\tau)})$, where $\overline{P}_B = \mathbbm{1}_B -P_{B}$. The output ensemble  $\mathcal{E} (\rho_{AB}^{(\tau)}, \{P_{B,i}^{(2)}\}_i)= \{ p_{A,i}, \, \rho_{A,i}\}_{i}$ is composed by an uniform distribution (by definition of ME-POVM) and states in the following form
\begin{equation}\label{p1}
p_{A,1}=   \frac{1}{2} \trB{\left( \rho_B'^{(\tau)} +\rho_B''^{(\tau)}\right) P_B  } =\frac{1}{2} \, ,
\end{equation}
\begin{equation}\label{p2}
p_{A,2} = \frac{1}{2} \trB{\left( \rho_B'^{(\tau)} +\rho_B''^{(\tau)}\right)\overline{P}_B  } =\frac{1}{2} \, ,
\end{equation}
\begin{equation}
\rho_{A,1}=   \ket{0}\bra{0}_A \trB{ \rho_B'^{(\tau)}  P_B  } +\ket{1}\bra{1}_A \trB{ \rho_B''^{(\tau)}  P_B  }  \, ,
\end{equation} 
\begin{equation}\label{rho2}
\rho_{A,2}=  \ket{0}\bra{0}_A \trB{ \rho_B'^{(\tau)}  \overline{P}_B  } +\ket{1}\bra{1}_A \trB{ \rho_B''^{(\tau)} \overline{P}_B }  \, .
\end{equation}
Since $\mathcal{E} (\rho_{AB}^{(\tau)}, \{P_{B,i}^{(2)}\}_i)$ is an equiprobable ensemble of two states, $P_g( \mathcal{E} (\rho_{AB}^{(\tau)}, \{P_{B,i}^{(2)}\}_i))= (2+||\rho_{A,1} -  \rho_{A,2}||_1)/4$. Hence, with Eqs. (\ref{p1})-(\ref{rho2}), we can write it as
$$
||  \, \ketbra{0}{0}_A \trB{ \rho'^{(\tau)}_B \cdot \Delta P_B} + \ketbra{1}{1}_A \!\trB{ \rho''^{(\tau)}_B \cdot\Delta P_B } ||_1 =
$$
$$  
= | \trB{ \rho'^{(\tau)}_B \cdot \Delta P_B }| + | \trB{ \rho''^{(\tau)}_B\cdot \Delta P_B }| \, ,
$$
where $\Delta P_B = P_B - \overline{P}_B$. Hence
$$
|| \rho_{A,1} - \rho_{A,2} ||_1 = \max_{\pm} |  \trB{ ( \rho'^{(\tau)}_B \pm \rho_B''^{(\tau)}) \cdot \Delta P_B}|  \, .
$$
 Using Eq. (\ref{p1}) and Eq. (\ref{p2}) we see that
$
| \trB{ ( \rho'^{(\tau)}_B + \rho_B''^{(\tau)}) \Delta P_B }| = | \trB{ ( \rho'^{(\tau)}_B + \rho_B''^{(\tau)})  P_B} - \trB{ ( \rho'_B + \rho_B''^{(\tau)}) \cdot \overline{P}_B }| = 2| p_{A,1} - p_{A,2} | =0 \, .
$
Hence:
$
|| \rho_{A,1} - \rho_{A,2} ||_1  =  | \, \trB{ ( \rho'^{(\tau)}_B - \rho_B''^{(\tau)}) (2P_B - \mathbbm{1}_B )}| = 2 |  \trB{ ( \rho'^{(\tau)}_B - \rho_B''^{(\tau)}) P_B}| \, ,
$
from which follows that
\begin{equation}\label{maximB}
C_B^{(2)} ({\rho}_{AB}^{(\tau)} )=\max_{ \{P_{B,i}^{(2)} \}_{i} \in \Pi_B( {\rho}_{AB}^{(\tau)} ) }  \frac{| \trB{ ( \rho'^{(\tau)}_B - \rho_B''^{(\tau)}) \cdot P_B}| }{2}  .
\end{equation}
To compare $C_B^{(2)} ({\rho}_{AB}^{(\tau)} )$ with $C_A({\rho}_{AB}^{(\tau)})$, we write
$$
C_A({\rho}_{AB}^{(\tau)}) = P_g (  \{ \{ p_{A,1,2}=1/2 \}_i , \{ \rho_B'^{(\tau)} , \rho_B''^{(\tau)} \} \} ) -\frac{1}{2} =$$
$$=\max_{ \{P_{B,i} \}_{i}   } 	\frac{  \trB{ \rho_B'^{(\tau)} \cdot P_B + \rho_B''^{(\tau)} \cdot \overline{P}_B } }{2} -\frac{1}{2}= $$
$$ {= \max_{ \{P_{B,i} \}_{i}  }}    \frac{   \trB{(\rho'^{(\tau)}_B - \rho''^{(\tau)}_B) P_B } }{2}
= \max_{ \{P_{B,i} \}_{i}  }   \frac{ |\trB{(\rho'^{(\tau)}_B - \rho''^{(\tau)}_B) P_B }| }{2}  \, .
$$
 The only difference between  $C_B^{(2)}(\rho_{AB}^{(\tau)}) $ and $C_A( \rho_{AB}^{(\tau)}) $ is in the maximization procedure: in the former we maximize only over the 2-output ME-POVMs $\Pi_B({\rho}_{AB}^{(\tau)})$, while in the latter we can pick any 2-output POVM: $C_A({\rho}_{AB}^{(\tau)}(t)) \geq C_B^{(2)}({\rho}_{AB}^{(\tau)}(t))$ follows as a natural consequence.

\section{Proof that $C_B(\rho_{AB}^{(\tau)}) = C^{(2)}_B (\rho_{AB}^{(\tau)})$}\label{2enough}

In this Appendix, in contrast to Appendix \ref{CB2CA}, we consider the action of any ME-POVM over $B$ for $\rho_{AB}^{(\tau)}$.
We want to show  that for each  ME-POVM $\{P_{B,i}^{(n)}\}_{i}$ that we can consider in $C_B(\rho_{AB}^{(\tau)})$, where $i$ runs from 1 to $n>2$, we can  always find at least one 2-output ME-POVM acting on $B$, i.e. $\{P_{B,1}, P_{B,2}  \} \in \Pi_B(\rho_{AB}^{(\tau)}) $, that provides an ensemble with a higher value of $P_g (\cdot) $. We recall that, if $\mathcal{E}=\{ p_i,\rho_i\}_i$ is a generic ensemble of $n$ states defined on $S(\mathcal{H})$, where $\mathcal{H}$ is a generic finite dimensional Hilbert space, the guessing probability of $\mathcal{E}$ is
\begin{equation}\label{Pg}
P_g (\mathcal{E})\equiv \max_{ \left\{ P_i \right\}_i }\sum_{i=1}^n  p_i \tr{ \rho_i \cdot P_i  } \, ,
\end{equation}
where the maximization is performed over the space of the $n$-output POVMs $\{P_i\}_i$ on $S(\mathcal{H})$.
 Starting from a general $n$-output ME-POVM $\{P_{B,i}^{(n)}\}_{i}$, we construct the corresponding 2-output ME-POVM $\{P_{B,1}, P_{B,2}  \} \in \Pi_B(\rho_{AB}^{(\tau)})$ that accomplishes this task.

For every given $n$-output ME-POVM $\{ P_{B,i}^{(n)} \}_i$ for ${\rho}_{AB}^{(\tau)}$, we can generate an equiprobable ensemble of states (EES) of the form $\mathcal{E}( {\rho}_{AB}^{(\tau)} , \, \{P_{B,i}^{(n)} \}_i) = \{\{p_i=1/n\},\{\rho_{A,i}\} \}_i$. The guessing probability of this ensemble, which we denote by  $P_g^{(n)} = P_g(\mathcal{E} (\rho_{AB}^{(\tau)},\{ P_{B,i}^{(n)} \}_i))$, is
\begin{equation}\label{Pgneven}
P_g^{(n)} =\tr{{\rho}_{AB}^{(\tau)}\cdot \left( \sum_{i=1}^n \overline{P}_{A,i}^{(n)} \otimes P_{B,i}^{(n)}  \right)   } \, ,
\end{equation}
where  $\{ \overline{P}_{A,i}^{(n)}\}_i$ is a POVM that provides the maximum in  Eq. (\ref{Pg}). If $n$ is even we consider the following 2-output POVM
\begin{equation}\label{2outputeven}
P^{(2)}_{B,1} = \sum_{i\in E_1} P_{B,i}^{(n)} \, , \,\,\,
P^{(2)}_{B,2} = \sum_{i\in E_2} P_{B,i}^{(n)} \, ,
\end{equation}
where $E_1$ and $E_2$ are any two sets of $n/2$ indices such that $E_1 \cup E_2 = \{ 1,2, \dots , n\}$. This structure guarantees that Eq. (\ref{2outputeven}) is a 2-output ME-POVM for $\rho_{AB}^{(\tau)}$. We compare Eq. (\ref{Pgneven}) with the guessing probability of the output ensemble that we obtain applying Eq. (\ref{2outputeven}) on $\rho_{AB}^{(\tau)}$
$$
P_g^{(2)}=\max_{\{P_{A,i}\}_{i=1,2} }  \tr{{\rho}_{AB}^{(\tau)} \cdot \left(  \sum_{i=1}^2 P_{A,i} \otimes P^{(2)}_{B,i}  \right)  } \geq $$
\begin{equation}\label{Pg2x} \geq \tr{{\rho}_{AB}^{(\tau)} \cdot \left(  \sum_{i=1}^2  P^{(2)}_{A,i} \otimes {P}^{(2)}_{B,i}  \right)  } \, ,
\end{equation}
where the POVM $\{ P^{(2)}_{A,i}\}_i $ is defined by 
\begin{equation}\label{PAiPBi}
P^{(2)}_{A,1} = \sum_{i\in E_1} \overline P_{A,i}^{(n)} \, , \, \,\,
P^{(2)}_{A,2} = \sum_{i\in E_2} \overline{P}_{A,i}^{(n)} \, .
\end{equation}

$$
P_g^{(2)}\geq \tr{ {\rho}_{AB}^{(\tau)} \cdot \left(P^{(2)}_{A,1}  \otimes P^{(2)}_{B,1} + P^{(2)}_{A,2}  \otimes P^{(2)}_{B,2}  \right) } 
= $$
$$
= \tr{  {\rho}_{AB}^{(\tau)} \cdot\left( \sum_{i=1}^n \overline P_{A,i}^{(n)} \otimes  P_{B,i}^{(n)} + P_{AB}^{mix} \right) } = $$
\begin{equation}\label{Pgeven}
= P_g^{(n)} + \tr{ {\rho}_{AB}^{(\tau)}  \cdot P_{AB}^{mix}   }  \geq P_g^{(n)}  \, ,
\end{equation}
where $P_{AB}^{mix}$ is a sum of mixed terms of the form $\overline  P_{A,i}^{(n)} \otimes  P_{B,j}^{(n)}$ with $i\neq j$, and it provides a non-negative contribution. 

On the other hand, if $n$ is odd, we define
\begin{equation}\label{PBodd}
P^{(2)}_{B,k} =\frac{1}{2} P_{B,x}^{(n)}+ \sum_{i\in O_k^x} P_{B,i}^{(n)} \hspace{0.75cm} (k=1,2) 
\end{equation}
\begin{equation}\label{PAodd}
P^{(2)}_{A,k} = \frac{1}{2} \overline P_{A,x}^{(n)}+ \sum_{i\in O_k^x} \overline P_{A,i}^{(n)}   \hspace{0.75cm} (k=1,2) 
\end{equation}
where $O_1^x$ and $O_2^x$ are any two sets of $(n-1)/2$ indices such that $O_1^x \cup O_2^x = \{ 1,2, \dots , n\} \setminus \! x $ (the value of $x$ will be fixed later). 
We consider again Eq. (\ref{Pg2x}), where $\{P^{(2)}_{B,i}\}_i$ is now given by Eq. (\ref{PBodd}) and and $P^{(2)}_{A,i}$ is now given by Eq. (\ref{PAodd}). Since $P^{(2)}_{A,i}$ is not necessarily a POVM that maximizes Eq. (\ref{Pg}) we have the following inequality for $P_g^{(2)}$
\begin{widetext}
$$
P^{(2)}_g \geq  \tr{ {\rho}_{AB}^{(\tau)}\cdot  \left(  \sum_{i\neq x} \overline P_{A,i}^{(n)} \otimes  P_{B,i}^{(n)} \,+ \frac{1}{2} \overline P_{A,x}^{(n)} \otimes P_{B,x}^{(n)}  + \frac{1}{2} \left(\sum_{i\neq x} \overline P_{A,i}^{(n)}  \right) \otimes P_{B,x}^{(n)}  + P_{AB}^{mix} \right)   } \geq 
$$
$$
=\tr{ {\rho}_{AB}^{(\tau)}  \cdot \left(  \sum_{i =1}^n P_{A,i}^{(n)} \otimes  P_{B,i}^{(n)} \,- \frac{1}{2} \overline P_{A,x}^{(n)} \otimes P_{B,x}^{(n)}  + \frac{1}{2} \left(\sum_{i\neq x} P_{A,i}^{(n)}  \right) \otimes P_{B,x}^{(n)}   \right)   } = 
$$
$$ 
= P_g^{(n)}  +  \tr{ \rho_{AB}^{(\tau)} \cdot \left( \frac{ - \overline P_{A,x}^{(n)} }{2} \otimes P_{B,x}^{(n)}  +  \frac{ \sum_{i\neq x} P_{A,i}^{(n)}  }{2}\otimes P_{B,x}^{(n)}   \right) }
= P_g^{(n)}  + \tr{\rho_{AB}^{(\tau)} \cdot \frac{ \mathbbm{1}_A - 2 \overline P_{A,x}^{(n)} }{2}  \otimes P_{B,x}^{(n)} }  ,
$$
\end{widetext}
where $P_{AB}^{mix}$ represents terms that provide positive contributions  to $P_g^{(2)}$. We have to find a value of $x$ that makes the second term of the last relation positive. Let $a_x$ and $b_x$ be the diagonal elements of $\overline P_{A,x}^{(n)}$ in the orthonormal basis $\{\ket{0}_A,\ket{1}_A\}$. We recall that $\rho_{AB}^{(\tau)} = ( \ketbra{0}{0}_A \otimes \rho_B'^{(\tau)}+ \ketbra{1}{1}_A \otimes \rho_B''^{(\tau)} )/2$
and we obtain
\begin{equation}\label{Pgodd}
P_g^{(2)} \!\geq \! P_g^{(n)}+  \mbox{Tr}_B{\left[ \!\left( \!\frac{1-2 a_x}{4} {\rho_B'}^{(\tau)} \!+\! \frac{1-2b_x}{4} \rho_B''^{(\tau)} \! \right)\cdot P^{(n)}_{B,x} \right]\!,}
\end{equation}
where the second term on the right-hand side of the inequality is definitely positive when $a_x,\, b_x \leq 1/2$. From  $\sum_i \overline P_{A,i}^{(n)}  = \mathbbm{1}_A $ follows that $\sum_{i=1}^n a_i = 1$ and $\sum_{i=1}^n b_i=1$. Therefore, if $a_x > 1/2 $ ($b_x > 1/2$), then $a_y \leq 1/2$ ($b_y \leq 1/2$) for any $y \neq x$. In order to fix the value of $x$, we must consider that $a_x$ and $b_x$ could be bigger than $1/2$ for two different values of $x$: let's say $x_a$ and $x_b$. Even in this ``worst-case'' scenario we still have $n-2$ other possible choices for $x$ such that $(1-2a_x),\, (1-2b_x) \geq 0$. We pick one of these values, and we call it $\overline x \in \{1,\dots,n \} \setminus \{x_a,x_b\}$. Finally, if we use ${\overline x}$ in the definition of the POVMs $\{ P^{(2)}_{A,i} \}_i$ and $\{ P^{(2)}_{B,i}\}_i$, from Eq. (\ref{Pgodd}) we obtain
\begin{equation}\label{Pgodd2}  
P_g^{(2)} \geq  P_g^{(n)}   \, .
\end{equation}

Equations (\ref{Pgeven}) and (\ref{Pgodd2}) show that, when we evaluate $C_B( \rho_{AB}^{(\tau)})$, the guessing probability of the ensembles generated by the $n$-output ME-POVMs is never bigger than the one that we obtain if we only consider the 2-output ME-POVMs:
$
C^{(2)}_B(\rho_{AB}^{(\tau)})= C_B(\rho_{AB}^{(\tau)}) \, .
$
Thanks to this result we can finally say that $C_A(\rho_{AB}^{(\tau)}(t)) \geq C_B(\rho_{AB}^{(\tau)}(t))$ and $C (\rho_{AB}^{(\tau)}(t))= C_A(\rho_{AB}^{(\tau)}(t))$. This result is valid if we consider $ \rho_{AB}^{(\tau)}$, but in general it is not true.

\section{Proof that $C_A({\rho}_{AB}^{(\tau)}) = C_A^{(2)} ({\rho}_{AB}^{(\tau)}) $}\label{2enoughA}

When we considered $C_A(\rho_{AB}^{(\tau)})$, we have seen that if the maximization over the ME-POVMs is considered only over the 2-output ones, the maximum is obtained for $\{P^{proj}_{A,i}\}_i=\{ \ketbra{0}{0}_A, \ketbra{1}{1}_A \}$. 
In order to complete the proof, we need to show that even if we consider general $n$-output ME-POVMs  (as in the definition (\ref{CAapp})), we don't get higher guessing probabilities of the corresponding output ensembles. In other words, if we use the definition
$$
C_A^{(2)} ({\rho}_{AB}^{(\tau)}) =\!\!\!\!\!\!\!\! \max_{ \left\{P_{A,i}\right\}_i \in \Pi_A^{(2)} \left( \rho_{AB}^{(\tau)} \right) } \!\!\!\!P_g \left( \mathcal{E} \left( \rho_{AB}^{(\tau)} ,\left\{ P_{A,i} \right\}_i \right)  \right)-\frac{1}{2} \, ,
$$
where $\Pi_A^{(2)}(\rho_{AB}^{(\tau)})$ contains only the 2-output ME-POVMs of $\rho_{AB}^{(\tau)}$, then $C_A(\rho_{AB}^{(\tau)}) = C_A^{(2)} ({\rho}_{AB}^{(\tau)}) $. 

To see this we can make the same analysis as done in Appendix \ref{2enough} for $C_B(\rho_{AB}^{(\tau)})$ but we switch the role of $A$ and $B$ in Eq. (\ref{2outputeven}) and Eq. (\ref{PAiPBi}) when $n$ is even and Eq. (\ref{PBodd}) and Eq. (\ref{PAodd}) when $n$ is odd. The definitions for $P_g^{(n)}$, $P_g^{(2)}$, $E_{1,2}$ and $O_{1,2}^x$ are preserved.

The guessing probability of an EES generated by a ME-POVM $ \{ P_{A,i}^{(n)}\}_i$ with an even number of outputs is
$$
P_g^{(n)} =\tr{{\rho}_{AB}^{(\tau)}\cdot \left( \sum_{i=1}^n {P}_{A,i}^{(n)} \otimes \overline P_{B,i}^{(n)}  \right)   } \, ,
$$
where $\{ \overline P_{B,i}^{(n)}\}_i$ is a POVM that maximizes the guessing probability in Eq. (\ref{Pg}).
The 2-output ME-POVM that provides a higher guessing probability is
\begin{equation}\label{PAiPBiA}
P^{(2)}_{A,1} = \sum_{i\in E_1}  P_{A,i}^{(n)} \, , \, \,\,
P^{(2)}_{A,2} = \sum_{i\in E_2} {P}_{A,i}^{(n)} \, .
\end{equation}
We define the following POVM on the system $B$
\begin{equation}\label{2outputevenA}
P^{(2)}_{B,1} = \sum_{i\in E_1} \overline P_{B,i}^{(n)} \, , \,\,\,
P^{(2)}_{B,2} = \sum_{i\in E_2} \overline P_{B,i}^{(n)} \, .
\end{equation}
Consequently, we consider the following inequality
\begin{eqnarray*}
&P_g^{(2)} \geq \tr{ {\rho}^{(\tau)}_{AB} \cdot \sum_{i=1,2} P^{(2)}_{A,i}  \otimes P^{(2)}_{B,i} } =  \\ 
&= P_g^{(n)}+ \tr{ {\rho}^{(\tau)}_{AB}  \cdot \sum_{k=1}^2 \sum_{i \neq j}^{  i,j\in E_k}    P_{A,i}^{(n)} \otimes  \overline P_{B,j}^{(n)}   } ,& 
\end{eqnarray*}
which shows that $P_g^{(2)} \geq  P_g^{(n)}$. 
If $n$ is odd, we use again the technique from Appendix \ref{2enough}, where we switch the role of $A$ and $B$, to obtain the inequality
$$
P_g^{(2)}\geq  P_g^{(n)}  + \tr{ \rho_{AB}^{(\tau)} \cdot \frac{ \mathbbm{1}_A - 2 P_{A,x}^{(n)} }{2}  \otimes \overline P_{B,x}^{(n)} }  ,
$$
where the right-hand side is greater than $P_g^{(n)}$ if $x$ is suitably chosen.
 
{
We underline that the results given in this section and Appendix \ref{CB2CA} suffice to state that $C^{(2)} (\rho^{(\tau)}_{AB})=C^{(2)}_A (\rho^{(\tau)}_{AB})\geq C^{(2)}_B (\rho^{(\tau)}_{AB})$.
}

\end{document}